# Autonomous Repair in Cementitous Material by Combination of Superabsorbent Polymers and Polypropylene Fibres: A Step Towards Sustainable Infrastructure

Souradeep GUPTA[a], KUA Harn-wei[b], PANG Sze-dai[c]

[a] *National University of Singapore, Singapore, souradeepnus@gmail.com*
[b] *National University of Singapore, Singapore, bdgkuahw@nus.edu.sg*
[c] *National University of Singapore, Singapore, ceepsd@nus.edu.sg*

## ABSTRACT

Manual maintenance and repair of cracks in concrete structures are often unsustainable because of associated labor, capital and environmental damage. Introduction of microfibers and superabsorbent polymers is a material solution to restrict crack propagation and enhance self-healing efficiency. Therefore, the study focus on development of a more sustainable cementitious system (cement mortar) comprising of polypropylene(PP) fibers and superabsorbent polymers (SAP) which would facilitate autonomous healing and recover original mechanical and durability properties of mortar.

To investigate crack sealing and recovery in mechanical and durability properties of mortar post healing, specimens were damaged by compressive and flexural loading. Mechanical strength, sorptivity and water penetration of healed mortars were compared to that of undamaged mortar at same age to estimate recovery of original properties while crack sealing was investigated by means of optical microscopy. Experimental results show mortar with combination of PP fibers and SAP showed full recovery of mechanical strength after healing while recovery in durability up to 90% was recorded. Microscopic images show that average crack-sealing ratio of 85% could be achieved under moist condition by combination of SAP and PP fibers while sealing under drier air curing condition is also significantly higher than reference samples with only fibers. Crack width up to 330μm has been found to be completely sealed by carbonate crystals. Furthermore, mortar with SAP and PP fibers retain about 70% of their original 28-day strength after three cycles of loading while reference mortar samples were found to retain only about 40-50% of their original strength.

Effective crack sealing and high recovery of original properties in mortars with SAP and fibers suggest that this material combination would reduce the need for environmentally damaging and expensive repairs during the service life which will be an important step towards achieving economical and environmentally sustainable construction practice.

## 1. INTRODUCTION

Cementitous materials are prone to cracking due to action of various loading and environmental factors. Cracks give easy passage to foreign chemicals and moisture into concrete structures which affect its serviceability. Therefore, if micro-cracks can be healed at early stage, it will reduce further propagation and arrest access of contaminants into the structure. Manual repair of cracks is often limited due to accessibility and high cost. Moreover, chemical or cement based repair materials present threats including material incompatibility, health and environmental hazards (De Muynck et al.,2010) which makes their use highly unsustainable. Self-healing in cementitious material is a sustainable solution to maintenance of concrete structures because it reduces the manual repair operations and concomitant environmental hazards. Although cementitious materials including cement mortar possesses autogenous self-healing capability by secondary hydration of calcium silicate hydrate and precipitation of calcium carbonate, the effectiveness of such healing mechanism is only limited to narrow cracks (Edvardsen, 1999; Ter Heide,2005). Yang et al. (2009) observed complete healing of crack width below 50 μm while cracks beyond 150μm were partially healed. It is concluded from the existing studies that the most essential conditions for effective autogenous self-healing in mortar include presence of calcium or carbonate ions, presence of moisture and restriction of crack width typically between 50 – 150 μm (Yang et al., 2009). Therefore, it means that effective healing can be obtained by a dedicated material design which can restrict crack width and source for moisture at the crack site for secondary hydration and precipitation of carbonate to seal cracks.

This study investigates the material design by combination of two different mechanisms on autogenous healing capacity in mortar. The first mechanism is the introduction of polypropylene (PP) fiber to restrict crack widths through bridging action during matrix micro-cracking or macro-cracking. The second mechanism involves the use of superabsorbent polymers (SAP) which can absorb high volume of moisture (up to 500 times of its own weight) and retain in its structure without dissolving. The moisture retained by SAP particles are gradually desorbed and made available for further hydration in the cement matrix. Lee et al. (2010) observed the healing effect of SAP by expansion and blocking of crack opening when moisture ingress through cracks takes place. SAP can also provide internal healing because it absorbs water from surrounding and seal cracks by secondary hydration. Snoeck et al. (2014) reported that combination of PVA fibers and SAP offer effective self-healing even when moisture is not available through secondary hydration and precipitation of calcium carbonate which seal cracks. However, besides sealing of cracks recovery of original properties post-healing is essential for delivering durable and sustainable infrastructure.

The objective of this study is to investigate the ability of combination of SAP and PP fibers to recover mechanical and durability properties of cementitious mortar after healing of damage created by structural loading. In addition, recovery of compressive strength after multiple cycles of loading and sealing of cracks under two different curing conditions is also studied in this article.

## 2. MATERIALS AND METHODS

The materials used in the study include CEMI 52.5 N Portland cement (500 kg/m$^3$) which meets ASTM C150 specification, natural sand (1375kg/m$^3$) with specific gravity of 2.55 and fineness modulus 2.54, polypropylene fibers (manufactured by W.R.Grace Singapore) and potassium based ionic superabsorbent polymers with average particle size ranging between 130-180 µm and water uptake capacity of 56±1g/g of SAP. The average length of polypropylene fibers used in the study is 19 mm with elastic modulus of 3.50 GPa and density 0.91g/cc.

The mix designs of different batches prepared are mentioned in Table 1. Water-cement ratio is maintained at 0.40 for all the mixes. From initial tests on effect of dosage on mortar properties it was confirmed that 0.70% of SAP and 0.60% addition of PP fibers by weight of cement do not affect strength and permeability. Therefore, the combination of PP fibers and SAP comprised of 0.70% SAP and 0.60% PP fiber by cement weight. Two other mixes - with only 0.60% fiber and only 0.70% SAP were studied as reference for comparison of recovery of original concrete properties with those in samples with combination of SAP and PP fibers.

| Mix codes | Mix description | Mix ratio (cement:sand:water:SAP:fiber) | Curing condition |
|---|---|---|---|
| Plain mortar | Without SAP and fiber | 1:2.75:0.4 | 27±2 ˚C, RH>95% |
| Fib0.60_REF1 | Mix with 0.60% PP fiber | 1:2.75:0.4:0:0.006 | |
| SAP0.70_REF2 | Mix with 0.70% SAP | 1:2.75:0.4:0.007:0 | |
| M1 | Mix with 0.60% PP fiber and 0.70% SAP | 1:2.75:0.4:0.70:0.007:006 | |

*Table 1: Mix design and description used in this study*

### 2.1 Methodology to study self-healing

At age of 14 days the samples are subjected to compressive strength test. Strength test has been done to ascertain the compressive strength of the samples from each batch to determine the pre-loading load levels which are 50% and 70% of 14-day strength. After introduction of damage by pre-loading at 14 day, the samples are allowed to cure at 25±2°C and > 95% RH for three weeks for the healing to occur. After three weeks of curing, the samples have been tested for mechanical strength and durability properties to calculate recovery post-healing. At the time pre-loaded samples are tested for recovery of mechanical and durability properties, the samples are 35 day old (14 day of initial curing +21 days of healing period) from the day of casting. Therefore, to estimate the extent of recovery of original properties (if there were no damage), undamaged samples were tested for the same mechanical and durability properties after 35 days of curing. While compressive and flexural strength tests have been conducted to enumerate recovery in strength, sorptivity and depth of water penetration tests are conducted to estimate recovery of durability properties post-healing. Minimum of three samples is tested to attain statistically relevant results. While mechanical strength testing and water penetration was conducted following BS EN standards, sorptivity has been conducted following ASTM C1585. Recovery of mechanical and durability properties upon self-healing is calculated using the following expressions.

$$\text{Recovery in strength} = \frac{\text{Peak strength (MPa) after healing}}{\text{Peak strength (MPa) of undamaged samples at 35 day}}$$

$$\text{Recovery in penetration depth} = \frac{\text{Maximum penetration (mm) in undamaged sample at 35 day}}{\text{Maximum penetration depth (mm) after healing}}$$

$$\text{Recovery in sorptivity} = \frac{\text{Initial sorptivity } \left(\frac{mm}{\sqrt{s}}\right) \text{ of undamaged samples at 35 day}}{\text{Initial sorptivity } \left(\frac{mm}{\sqrt{s}}\right) \text{ of healed samples}}$$

Recovery of initial sorptivity is considered because initial sorptivity is more sensitive to presence of micro-cracks induced by preloading of the specimens. Secondary sorptivity is contributed mostly by air voids in the specimen and therefore excluded from the calculation of recovery in sorptivity. Recovery greater equal to 1 would mean that the healed samples have completely regained the original concrete properties while a number less than 1 means partial healing of samples in terms of recovery of strength or durability.

### 2.2 Recovery of compressive strength after multiple loading cycle

Recovery of compressive strength under multiple cycles of loading was studied by subjecting cylinder specimens to three cycles of loading. In each cycle the specimens were loaded to 70% of their ultimate strength and were allowed to cure for 21 days between subsequent cycles of loading. The recovery of compressive strength after three cycles is expressed as percentage of 28-day strength of the mix.

$$R(\%) = \frac{\text{Compressive strength (MPa) after three cycles of loading}}{\text{28 day compressive strength of the mix}} * 100$$

### 2.3 Measurement of crack sealing

Sealing of cracks was monitored and recorded over a period of three weeks using optical microscope (Olympus SZX10 with Leica illumination system). Prism samples (40x40x160mm) were unloaded at the point of flexural failure and width of the crack formed at failure has been recorded. For controlled creation of crack steel fibers were placed in prism samples targeted for study of crack sealing. Cracked samples were moist cured except M1 which was both moist and air-cured during the healing period. The percentage reduction in crack width is calculated by using the following expression.

$$\text{Percentage reduction in crack width}(\%) = \frac{\text{crack width}_i - \text{crack width}_t}{\text{crack width}_i} \times 100$$

crack width$_t$ refers to average crack width recorded at time t (for example, t = 5 when crack width is measured on day 5).

crack width$_t$ refers to average initial crack width recorded just after introduction of cracks.

## 3. RESULTS AND ANALYSIS

### 3.1 Recovery of compressive strength

It can be observed from Figure 1 that recovery ratio of strength is considerably higher in M1 at 50% and 70% preloading compared to the reference although 35-day compressive strength of M1 is similar to that of pain mortar and SAP0.70_REF2.

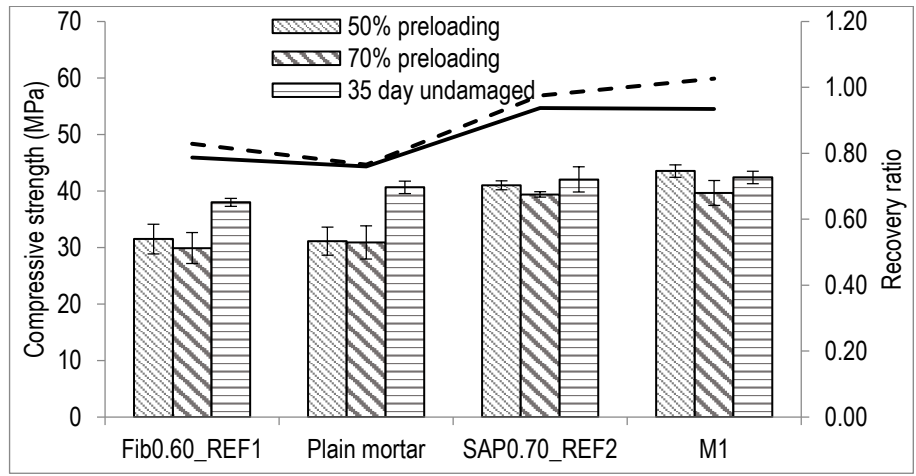

*Figure 1: Compressive strength of healed samples, 35day undamaged samples and recovery ratio at 50% and 70% preloading*

It can be explained by joint action of fiber and SAP where fibers tend to restrict crack width and SAP particles improve the rate of autogenous healing by supplying moisture to interior of mortar. Generation of hydration products bridge the cracks and impart strength to damaged mortar post healing. Moreover, water desorbed from SAP is rich in calcium ion which reacts with carbon dioxide from surrounding environment resulting in sealing of cracks by precipitation of calcium carbonate at the crack face. Apart from restricting crack width and increasing efficiency of autogenous healing, fibers act as anchors for calcium carbonate crystals. In absence of fibers, the carbonate crystals may not have any anchor to attach to and may be washed away by incoming fluid. SAP0.70_REF2 contained SAP but absence of fibers had little effect on restricting crack propagation which explains its lower compressive strength post-healing.

### 3.2 Recovery of water penetration

Figure 2 shows samples with SAP and fiber shows higher recovery ratio upon healing. In case of plain mortar and fiber reinforced mortar, healing of microcracks take longer time because of unavailability of moisture in the interior of mortar resulting in partial healing of cracks. Therefore, water under pressure can flow into the concrete without much resistance. In SAP containing samples, healing is more complete because SAP particles can absorb moisture from the moist environment (during curing) and supply for hydration to continue and heal the cracks. Moreover, near the crack faces where there are hydration products calcium carbonate precipitation takes place due to presence of calcium ion (Snoeck and De Belie, 2015).

2869

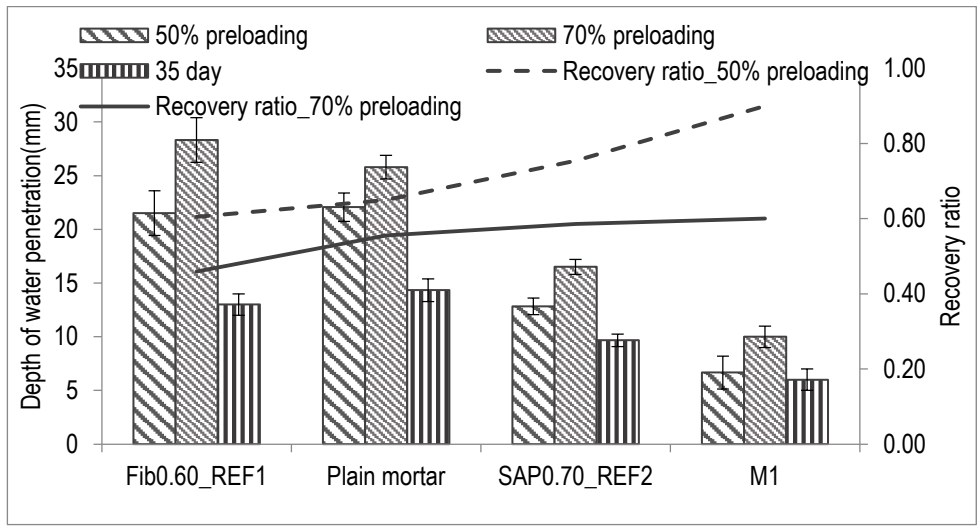

*Figure 2: Water penetration depth in healed samples, 35day undamaged samples and recovery ratio at 50% and 70% preloading*

Formation of calcium carbonate bridges and seals the cracks. In presence of fibers across microcracks, calcium carbonate can attach to fibers which prevent them from being washed off from the crack face. This explains higher healing ratio in M1 samples. Moreover, voids containing SAP particles create weak planes which make cracks to cross those voids. SAP s along crack faces swell immediately when it comes in contact with water thus blocking further penetration (Snoeck et al., 2014). However, when preloaded to high stress level localized wider cracks are formed which may not be sufficiently bridged by fibers or blocked by swelled SAP particles. It explains similar recovery ratio in M1 and SAP containing reference samples at 70% preloading level.

### 3.3 Recovery of sorptivity

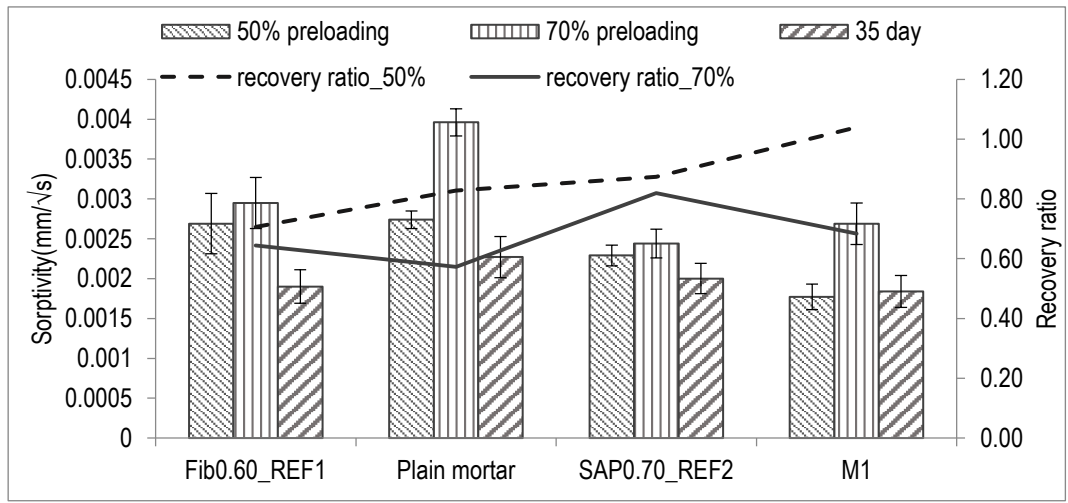

*Figure 3: Sorptivity in healed samples, 35day undamaged samples and recovery ratio at 50% and 70% preloading*

From Figure 3, one can observe that recovery of sorptivity is highest for M1 samples when subject to 50% preloading although at 70%, the recovery ratio is similar to the reference mortars. This may be attributable to crack profiles created in fiber reinforced mortar under stress. Incorporation of fibers bridge cracks and reduce permeability when the loading is below a certain limit called the threshold level. Above this threshold fibers may cause more localized unrecoverable deformation compared to plain mortar (Hosseini et al., 2009). 70% preloading may have introduced local deformation in the mortar that could not be fully healed or recovered which explains higher sorptivity in M1 samples at 70% preloading.

### 3.4 Recovery of compressive strength under cyclical loading

From Figure 4 one can observe that mix M1 retains highest percentage, about 70% of 28-day compressive strength after being subject to three loading cycles.

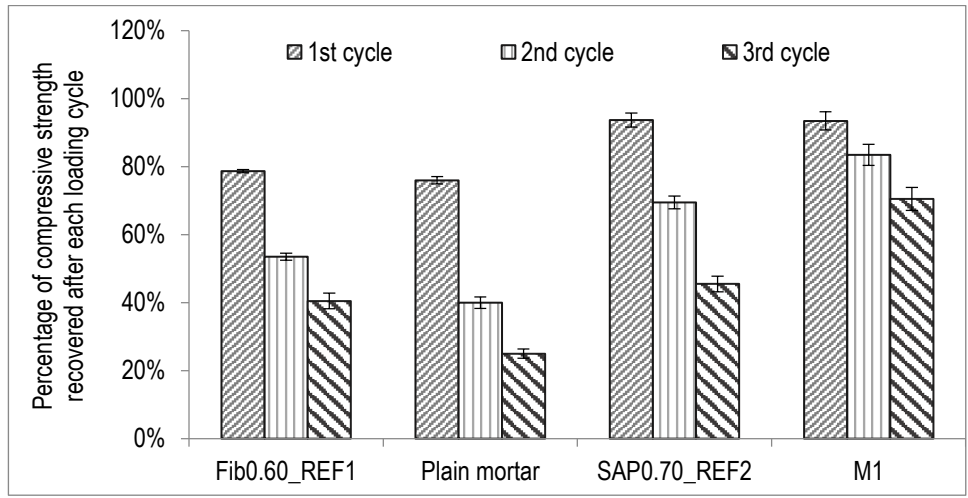

*Figure 4: Recovery percentage of compressive strength of mixes subject to three cycles of loading*

It may be justified by the role played by PP fibers in mitigating crack propagation which is also observed when Fib0.6_REF 1 and plain mortar are compared. Plain mortar regained only about 25% of its 28-day strength after three loading cycles compared to 40% in case of Fib0.6_REF1. Addition of SAP also plays an important part in accelerating the healing process. With age the microstructure of mortar becomes denser and therefore less external water tends to reach interior of concrete. Therefore, autogenous healing is mitigated due to lower availability of water. Due to their affinity to water, SAP plays an important role by uptaking moisture from the environment and supplying them for faster autogenous healing in damaged mortar. Therefore, controlled crack width due to addition of fibers and autogenous healing by SAP better performance of M1 samples under cyclical loading.

### 3.5 Sealing of cracks

Figure 5 shows that samples with SAP and fiber showed about 85% crack healing compared to only 30-40% healing in reference samples. Figure 6 (b) and (c) shows only partial healing in plain mortar and Fib0.6_REF1 samples which is attributed to limited availability of moisture at the crack site in absence of SAP particles.

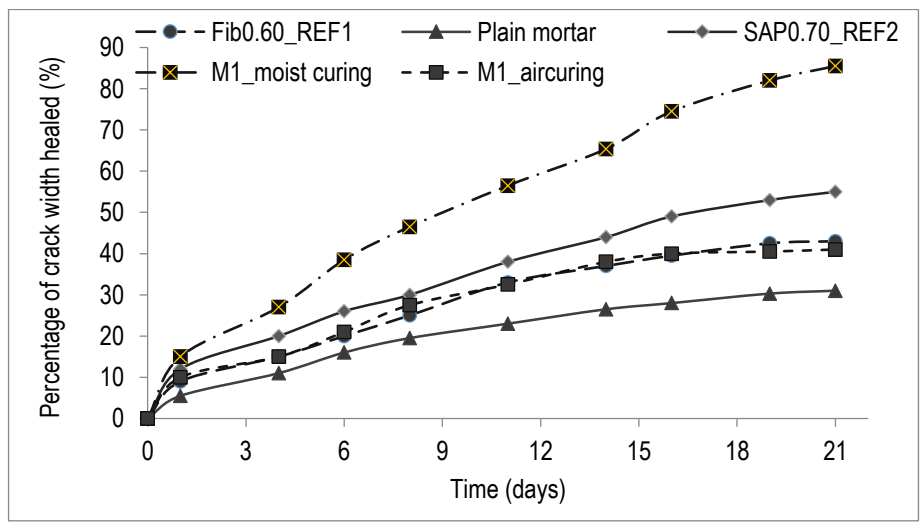

*Figure 5: Percentage of crack width healed over the healing period in different mixes under moist curing condition and in M1 under moist curing and air curing condition*

2871

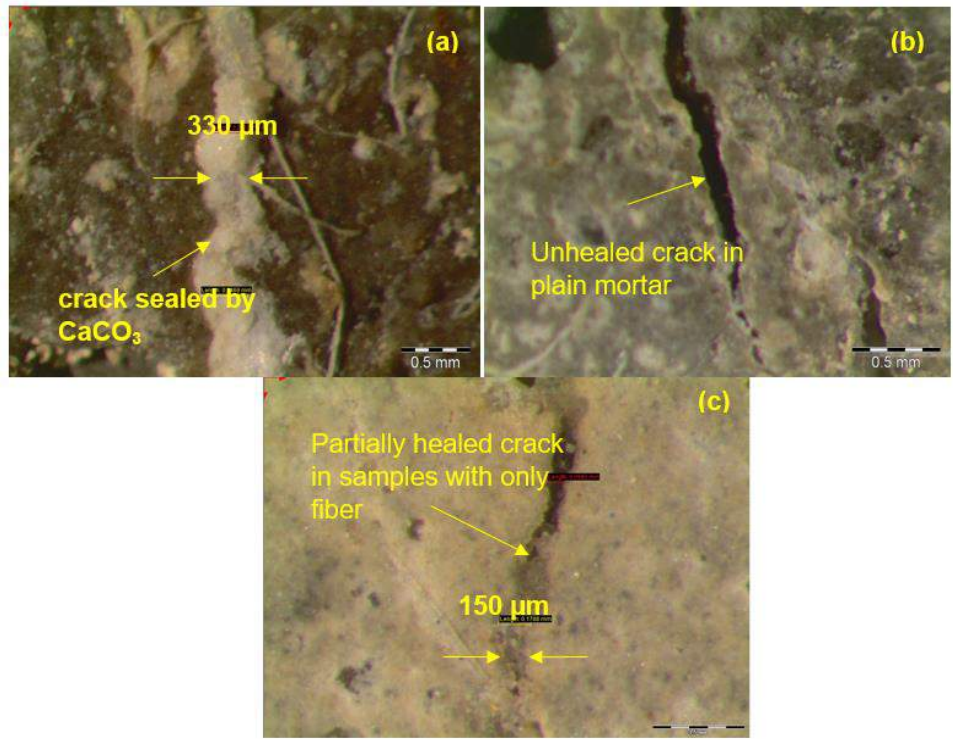

*Figure 6: Crack healing in (a)M1 sample (b)Plain mortar samples and (c)Fib0.6_REF1 under moist curing condition*

The range of crack width created in M1 samples was 220 – 380 µm at failure while it was in the range of 200 - 300 µm in Fib0.6_REF1. It means that M1 showed slightly higher crack width at failure which may be attributed to weak spots created due to desorption of water from SAP. While complete sealing of cracks up to 330µm was noted in M1 (Figure 6a), cracks beyond 330µm were healed partially.

The unhydrated cement particles near the crack location had ready access to moisture supplied by SAP for further hydration. Moreover, calcium ions at crack face may have reacted with water containing dissolved carbon dioxide or bicarbonate to form calcium carbonate which sealed the cracks. M1samples placed under air curing also shows similar healing ratio as Fib0.60_REF1 which suggests that presence of SAP particles ensure availability of moisture even under dry condition to seal cracks by further hydration or precipitation of calcium carbonate. However, rate of healing under air-curing dropped beyond 12 days because most of the water absorbed by SAP has been desorbed and utilized for hydration.

## 4. CONCLUSION

The major findings from this study can be summarized as follows:

- Almost complete recovery of mechanical strength has been observed for samples containing SAP and fibers which is considerably higher than samples containing only fibers. It means that durable and strong infrastructure can be delivered by combining reinforcing effect of fibers and healing effect of SAP particles.
- SAP particles play an important role in sealing of crack width. In three weeks, about 85% crack width healing was observed in samples containing SAP and fibers of carbonate and formation of hydration products to seal up the cracks.
- Penetration of moisture is significantly reduced by action of SAP and fibers in healed and undamaged samples which mean that healed structures would be much less prone to damage by ingress of foreign chemicals and would require lower maintenance.

The findings suggest that SAP containing self-healing system would be most suitable in regions with high humidity. The SAP s can draw moisture from the air to facilitate self-healing by further hydration and carbonate precipitation. However, further research is to be conducted to study this system's effectiveness in dry climate because lack of moisture may slow down or affect the rate of autonomous healing.

Although superabsorbent polymers do not have any health hazard, some precautions may be needed while handling of polypropylene fibers. Fine polypropylene fiber filaments may penetrate in to respiratory system by inhalation that creates respiratory symptoms. Therefore, use of personal protection equipment (PPE), especially face mask and gloves are recommended while handling polypropylene fibers during preparation of concrete mix.